\documentclass{emulateapj} 
\usepackage{graphicx}
\usepackage{color,calc}
\usepackage{amsmath}
\usepackage{amssymb}
\usepackage{amstext}

\begin{document}

\title{X-Ray Binaries and the Current Dynamical States of Galactic Globular Clusters}
\shorttitle{X-Ray Binaries and the Current Dynamical States of Galactic Globular Clusters}
\submitted{ApJL, accepted} 
\author{John M. Fregeau\altaffilmark{1,2}}
\shortauthors{FREGEAU}
\affil{Department of Physics and Astronomy, Northwestern University, Evanston, IL 60208}
\altaffiltext{1}{fregeau@northwestern.edu}
\altaffiltext{2}{Chandra Fellow}

\begin{abstract}
It has been known for over 30 years that Galactic globular clusters (GCs) are overabundant
by orders of magnitude in bright X-ray sources per unit mass relative to the disk
population.  Recently a quantitative understanding of this phenomenon has developed,
with a clear correlation between the number of X-ray sources in a cluster, $N_X$, and the cluster's
encounter frequency, $\Gamma$, becoming apparent.  We derive a refined version
of $\Gamma$ that incorporates the finite lifetime of X-ray sources 
and the dynamical evolution of clusters.  With it we find we are able
to explain the few clusters that lie off the $N_X$--$\Gamma$ correlation,
and resolve the discrepancy between observed GC core radii
and the values predicted by theory.  
Our results suggest that most GCs are still in the
process of core contraction and have not yet reached the thermal equilibrium
phase driven by binary scattering interactions.
\end{abstract}

\keywords{globular clusters: general --- methods: numerical --- stellar dynamics}

\section{X-Ray Sources and Cluster Dynamics}\label{sec:dyn}
It was realized more than 30 years ago that Galactic globular clusters (GCs) are overabundant
by orders of magnitude in bright X-ray sources per unit mass relative to the 
disk population \citep{1975ApJ...199L.143C,1975Natur.253..698K}.
It was quickly understood that strong dynamical scattering interactions of
binaries in the dense cluster cores should be responsible for this overabundance
\citep{1987IAUS..125..187V}.  With the advances in X-ray astronomy
made possible by observatories such as {\em Chandra}, the relationship
between X-ray sources and core cluster dynamics has recently
been quantitatively studied.  \citet{2003ApJ...591L.131P} performed {\em Chandra}
observations of many Galactic GCs down to a limiting luminosity
of $4 \times 10^{30}\,{\rm erg}/{\rm s}$ in the 0.5--6 keV range (which includes
low-mass X-ray binaries [LMXBs] in outburst and quiescence, cataclysmic variables [CVs], millisecond
pulsars [MSPs], and magnetically active main sequence binaries [ABs]), and looked
for correlations between the number of X-ray sources in each cluster and 
properties of the cluster itself.  They found the strongest correlation with
the ``encounter frequency'' $\Gamma$, a rough estimate of the current
dynamical encounter rate in the cluster.  More recently, \citet{2003ApJ...598..501H}
and \citet{2006ApJ...646L.143P} have isolated the quiescent LMXBs (qLMXBs) 
and CVs, respectively, from the X-ray source populations, and have shown that their numbers are
indeed consistent with dynamical formation.

These results represent quantitative, 
empirical evidence that dynamical encounters are responsible for the formation
of X-ray sources in clusters.  However, they suffer from at least a few drawbacks.
First, the correlation between the number of X-ray sources, $N_X$, and the
encounter frequency appears to be sub-linear, with $N_X \propto \Gamma^{0.74 \pm 0.36}$,
although for LMXBs the exponent is $0.97 \pm 0.5$ \citep{2003ApJ...591L.131P}.  
Second, there are three clusters for which $N_X$
is significantly larger than predicted by $\Gamma$.  In the original
work of \citet{2003ApJ...591L.131P} it was already clear that
NGC 6397 has an $N_X$ that is $\sim 5$ times larger than predicted by the $N_X$--$\Gamma$ correlation.
Recent observations show that $N_X$ is factor of $\sim 2$ 
times that predicted by $\Gamma$ for NGC 7099 \citep{2007ApJ...657..286L},
and a factor of $\sim 20$ for Ter 1 \citep{2006MNRAS.369..407C}.
The common thread among these three clusters is that they are observationally
``core-collapsed,'' while all others in the \citet{2003ApJ...591L.131P} sample are not
\citep{2006AdSpR..38.2923G}.
(A possible exception is NGC 6752, whose collapsed core status is debated 
\citep{2003ApJ...595..179F,1995ApJ...439..191L}.)  A cluster is observationally
termed core-collapsed if its surface brightness profile is consistent with a cusp
at the limit of resolution, making it more difficult to measure
the core radius.   As described below, the collapsed core status of a cluster
is linked to its dynamical state, implying that cluster evolution
complicates the $N_X$--$\Gamma$ correlation.

\begin{figure}
  \begin{center}
    \includegraphics[width=0.9\columnwidth]{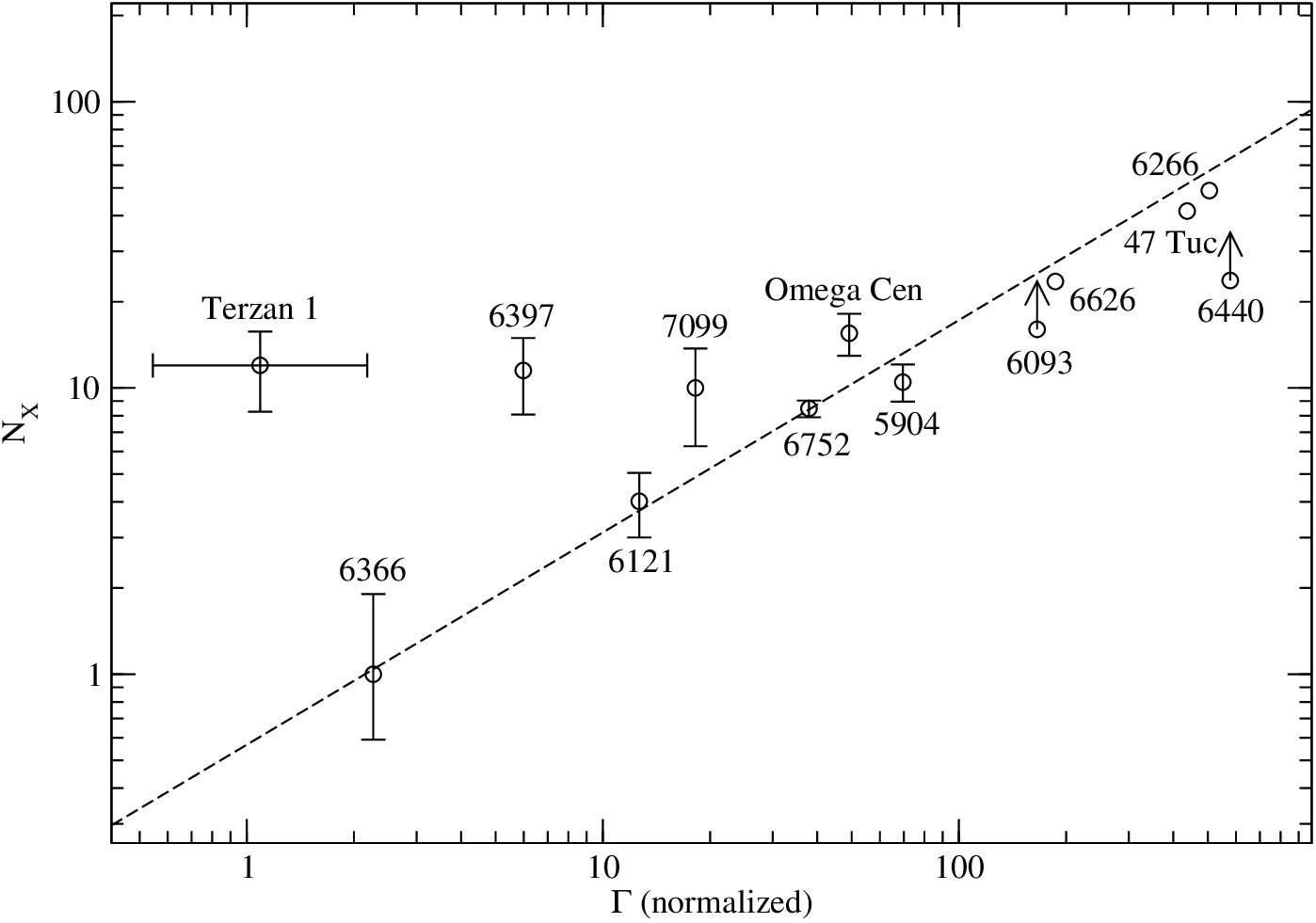}
    \caption{Number of observed cluster X-ray sources with 
      $L_X\gtrsim 4 \times 10^{30}\,{\rm erg}/{\rm s}$ for several Galactic
      GCs versus the encounter rate $\Gamma$.
      The power-law fit and data points are from \citet{2003ApJ...591L.131P} with the exception
      of NGC 7099 \citep{2007ApJ...657..286L} and Ter 1 \citep{2006MNRAS.369..407C}.
      The $N_X$ error bars for Ter 1, NGC 6397, and NGC 7099 represent source counting
      noise and background source uncertainty, but for the remaining
      clusters represent only background uncertainty.
      \label{fig:gamma}}
  \end{center}
\end{figure}

\section{Understanding Cluster Core Radii}\label{sec:rcrh}

The evolution of a GC, being a bound self-gravitating system, is very similar
to the evolution of a star, and comprises three main phases.
In the ``core-contraction phase,'' the first phase of evolution, 
the cluster's core contracts on a 
relaxation timescale, much like a pre-main sequence star.  Once the core density becomes
large enough for binary stars to begin strongly interacting 
dynamically, and thus generating energy via super-elastic encounters, the cluster settles into the
``binary-burning phase,'' analogous to the main sequence in stars.  In this phase the dynamical
properties of the cluster core remain roughly constant 
\citep[e.g.,][]{1991ApJ...370..567G,2003gmbp.book.....H,2007ApJ...658.1047F}.  
Once the binary
population is exhausted in the cluster core, it will collapse\footnote{Note that due to 
clashing naming conventions between observation and theory, this core collapse does not 
necessarily correspond to the observational definition, as discussed in the text.  We 
therefore suggest that some non-dynamical term be used in place of the observational 
``core-collapsed'' designation, or at least that the distinction between the observational
and theoretical uses of the term be noted when it is used.} 
via the gravothermal instability,
leading to extremely high central densities.  Deep in collapse, an energy producing event, 
such as an interaction of a dynamically-formed binary, will reverse the collapse, causing the core to
rebound and enter the ``gravothermal oscillation phase,'' in which the core continues to collapse
and rebound \citep{2003gmbp.book.....H}.  For a graphical representation of the three main phases of 
cluster evolution, see Figure 1 of \citet{1991ApJ...370..567G}, Figure 5 of \citet{2003ApJ...593..772F},
or Figure 29.1 of \citet{2003gmbp.book.....H}.

Since there is strong observational evidence that Galactic GCs were born with
significant binary fractions \citep{1992PASP..104..981H},
and since the binary-burning phase is the longest-lived phase of cluster evolution 
\citep[perhaps tens of Hubble times;][]{1991ApJ...370..567G,2003gmbp.book.....H,2007ApJ...658.1047F,2007MNRAS.374..344T}, 
it is widely
believed that most clusters observed today should be in this phase.  Early approximate 
calculations suggested that
the ratio of core to half-mass radius, $r_c/r_h$, in the binary burning phase, is broadly
consistent with observations of the $\sim 80$\% of Galactic GCs that are
{\em not} observationally core-collapsed \citep{1991ApJ...370..567G,2003ApJ...593..772F}.
Recently, however, more accurate simulations have shown that the early calculations overestimate
$r_c/r_h$ by a factor of 10 or more \citep{2006MNRAS.368..677H,2007ApJ...658.1047F}.
These latest results are quite difficult to ignore, since they represent the concordance
of two completely independent cluster evolution codes---one direct $N$-body, with minimal
approximations and a natural inclusion of binary interactions; the other the approximate 
H\'enon Monte Carlo method, with direct few-body 
integration of binary interactions.  The values of $r_c/r_h$
in the binary-burning phase agree quite well between the two codes, and furthermore agree quite well with 
semi-analytical theory \citep{1994ApJ...431..231V}.  The result is that now only 
the core-collapsed clusters agree with the predicted values of $r_c/r_h$ in the 
binary-burning phase, implying that if most clusters are in this phase 
some other energy generation mechanism is responsible for the measured core sizes.  
Several suggestions have been put forth for the energy
source, including, most notably, central intermediate-mass black holes (IMBHs) \citep{2006astro.ph.12040T}.

One key feature of the evolution of GCs that has escaped careful 
attention, though, is the timescale of the initial phase of core contraction.  
Many numerical simulations have shown that it is of order
$\sim 10$ {\em initial} half-mass relaxation times 
\citep{1991ApJ...370..567G,2003MNRAS.343..781G,2006MNRAS.368..677H,2007ApJ...658.1047F}.  
Since the 
current half-mass relaxation time for most clusters is $\sim 1$ Gyr and was likely much longer
in the past, the core contraction phase may easily last
longer than a Hubble time \citep[e.g.,][]{2007MNRAS.379...93H}.  
In fact, recent $N$-body simulations have shown that 
for a range of initial binary fractions, the evolution of $r_c/r_h$ over a Hubble time 
is a simple core contraction with no evidence of a binary-burning phase being reached
\citep{2007MNRAS.379...93H}.  Thus the solution to the discrepancy between theory and 
observations in $r_c/r_h$ could be the most mundane one, namely that core-collapsed
clusters are in the binary-burning phase while the rest are still undergoing core contraction.

\section{A Refined $\Gamma$}\label{sec:gamma}

First introduced by \citet{1987IAUS..125..187V}, the encounter frequency
$\Gamma$ is an estimate of the {\em current} dynamical interaction rate
in the cluster, which is assumed to be proportional to the current
number of observable X-ray sources.  We refine this predictive quantity by including, among
other factors, the recent history of the evolution of the core properties.
We start with the general form of the interaction
rate, then specialize to the \citet{1987IAUS..125..187V} $\Gamma$, 
and our new version.

The interaction rate between two species of objects can be written generally as
\begin{equation}\label{eq:main}
  \Gamma \equiv \frac{dN_{\rm int}}{dt} = \int\!\!\!\int n_1 n_2 \sigma_{12} \left| {\bf v}_{12} \right|
  f({\bf v}_{12}) d^3{\bf v}_{12}\,d^3{\bf r} \, ,
\end{equation}
where $n_i$ is the number density of species $i$, $\sigma_{12}$
is the interaction cross section between the two species, ${\bf v}_{12}$ is their relative
speed with $f$ its distribution function, 
and the integral is carried out over relative velocity
space and volume.  For the dynamical creation of LMXBs, species 1 represents stellar binaries,
while species 2 represents neutron stars.  The dynamical formation of CVs and other 
low-luminosity X-ray sources is rather more complicated, so species 1 and 2 generally represent single and
binary star systems \citep[see, e.g., Figure 7 of][]{2006MNRAS.372.1043I}.  
Typically, the integral in eq.~(\ref{eq:main}) is 
approximated as
\begin{equation}\label{eq:standardgamma}
  \frac{dN_{\rm int}}{dt} \propto \rho_c^2 r_c^3 / v_\sigma \, ,
\end{equation}
where $\rho_c$ is the core mass density, $v_\sigma$ is the core velocity dispersion,
the integral has been approximated by the core value, 
the gravitational-focusing dominated cross section has been used, 
and $\rho_{\rm c,1}/\rho_{\rm c,2}$ is assumed to be constant for all clusters.
Additionally, it should be pointed out that only the proportionality 
in eq.~(\ref{eq:main}) has been preserved since several factors (some of which are not 
constant among clusters) have been dropped.  More accurate approximations of eq.~(\ref{eq:main})
have been used, including numerical integrals over cluster models \citep{2003ApJ...591L.131P}, 
but all are estimates of the {\em current} interaction rate.

Dynamically formed X-ray binaries (XRBs) are known to have finite detectable lifetimes.  For LMXBs, 
this lifetime varies from $\sim 10^5$--$10^7$ yr for red giant donors, to $\sim 1$ Gyr for main-sequence
companions, to a few Gyr for ultracompacts \citep{2007arXiv0706.4096I}.  For CVs, this 
lifetime is $\sim 1\,{\rm Gyr}$ (N. Ivanova, priv.\ comm.).  An additional complication 
is that dynamically formed XRBs do not necessarily turn on as X-ray sources immediately following a
strong interaction.  In fact, the interaction that places a binary on the path to becoming
an observable X-ray source typically occurs several Gyr before it becomes detectable
\citep{2006MNRAS.372.1043I,2007arXiv0706.4096I}.

Since the XRBs we see now formed a few Gyr
or more ago, and since the recent dynamical history of clusters may have been quite variable,
it is clear that the current number of observable sources should be proportional to the
interaction rate integrated over time.  We thus write
\begin{equation}
  \Gamma \equiv N_{\rm int} = \int\!\!\!\int\!\!\!\int n_1 n_2 \sigma_{12} \left| {\bf v}_{12} \right|
  f({\bf v}_{12}) d^3{\bf v}_{12}\,d^3{\bf r}\,dt \, .
\end{equation}
We perform the time integration over an interval $t_x$, the typical detectable
lifetime of an XRB, to a time $t_\ell$, the typical timescale between
strong interaction and observability, in the past.
We leave $t_x$ and $t_\ell$ as parameters, but take $t_x=1\,{\rm Gyr}$ and
$t_\ell=3\,{\rm Gyr}$ as canonical values.
We further simplify the integral by writing
\begin{equation}
  N_{\rm int} = \int_{t_0-t_\ell-t_x}^{t_0-t_\ell} f_b f_{\rm co} n_c^2 \sigma_{\rm strong} \sqrt{2} v_\sigma
  \frac{4\pi}{3} r_c^3 \,dt \, ,
\end{equation}
with $f_b$ the core binary fraction, $f_{\rm co}$ the core compact object fraction, 
$n_c$ the core number density, $r_c$ the core radius, $v_\sigma$
the 1-D core velocity dispersion, and $\sigma_{\rm strong}$ the cross section for a strong
interaction between a binary and a single star.  The factor of $\sqrt{2}$ 
is from taking the difference of two Maxwellian velocity distributions.  
The cross section is
$\sigma_{\rm strong} \approx \pi a 2 G M / (\sqrt{2} v_\sigma)^2$, where $a$ is the
typical semi-major axis of a binary, $M=3m$ is the total mass of the binary--single
system, and $m$ is the typical stellar mass.  Taking $f_b$ and $f_{\rm co}$ to be 
constant over the time integral and substituting the
definition of the core radius, $r_c^2=9v_\sigma^2/4\pi G m n_c$
\citep{2003gmbp.book.....H}, yields
\begin{equation}\label{eq:81}
  N_{\rm int} = f_b f_{\rm co} \frac{81 \sqrt{2} a}{4 G m} 
  \int_{t_0-t_\ell-t_x}^{t_0-t_\ell} v_\sigma^3 r_c^{-1} \,dt \, .
\end{equation}
This expression is similar to eq.~(\ref{eq:standardgamma}), but with the core
properties integrated over time.
We have kept all numerical factors for the sake of completeness---only
the proportionality represented by this equation is needed for what follows.
By substituting in for the time evolution of the core quantities in the integrand,
one can use this expression to differentiate among the three different phases of cluster
evolution.  
We exclude the gravothermal oscillation phase, since in this phase a cluster
should have no more than a few binaries in its core \citep{2003gmbp.book.....H}, and all clusters
considered here with measured core binary fractions show evidence for many more
binaries than this 
\citep[note that a recent estimate puts the core binary fraction of NGC 6397 at $15\pm 1$\%;][]{davisposter}.

In the binary-burning phase the core radius and velocity dispersion are constant, so
the integral is easy to evaluate.  For the core contraction phase, we adopt the time
evolution of the core radius shown in the $N$-body models of \citet{2007MNRAS.379...93H},
approximated as
$r_c(t) = r_{c,0} (10-9t/t_0)$, with $r_{c,0}$ the current core 
radius and $t_0$ the current cluster age.  Since there is essentially
no core energy support in the core contraction phase, for the relationship between core
radius and central velocity dispersion we adopt the self-similar collapse model,
with $v_\sigma^2 \propto r_c^{-0.21}$ \citep{1987gady.book.....B}.  The ratio of the number
of interactions for a cluster in the binary-burning phase to the same cluster in core 
contraction is then
\begin{equation}
  \frac{N_{\rm int,bb}}{N_{\rm int,cc}} = \frac{t_x}{t_0}
  \frac{2.835}{\left(\frac{t_0}{t_0+9t_\ell}\right)^{0.315}-
    \left(\frac{t_0}{t_0+9t_\ell+9t_x}\right)^{0.315}} \, ,
\end{equation}
which has a minimum of $2.0$ and a maximum of $17.8$ in the range
$t_x=10^{-4}$--$3\,{\rm Gyr}$,
$t_\ell=1$--$10\,{\rm Gyr}$, for $t_0=13\,{\rm Gyr}$.
For the canonical values of $t_x=1\,{\rm Gyr}$
and $t_\ell=3\,{\rm Gyr}$ with $t_0=13\,{\rm Gyr}$, the value is $5.0$.  Since the 
number of X-ray sources should scale roughly linearly with the number of interactions,
this suggests that if a cluster is in the binary-burning phase (and has been for a time
$t_\ell+t_x$ to the present), it should have $\sim 5$ times
as many X-ray sources than it would if it were in the core contraction phase.
(If the \citet{2003ApJ...591L.131P} exponent of $0.74$ is adopted
this factor is $3.3$.)
Interestingly, the three clusters in Figure~\ref{fig:gamma} that
are observed to be core collapsed are
the three that have a significantly larger $N_X$ 
than predicted by the standard $N_X$--$\Gamma$ correlation, by a factor of 
$\sim 2$ to $\sim 20$.  NGC 6397 has an overabundance factor of
$\sim 2$, NGC 7099 a factor of $\sim 5$, and Ter 1 a factor of $\sim 20$.
However, we note that the estimate of $\Gamma$ for Ter 1 is more uncertain
than the rest since it comes directly from eq.~(\ref{eq:standardgamma}) and not
an integral over the cluster profile, and since the velocity dispersion
is only estimated as it has not been measured observationally.
The similarity between the observed overabundances and those predicted by our
simple revision of $\Gamma$
suggests that the observationally core-collapsed clusters are indeed in the
binary-burning phase, while the rest are still in the process of core contraction.

A natural objection to this conclusion is that cluster metallicity may explain
the X-ray source overabundance.  Studies of bright LMXBs associated with star
clusters in external galaxies have shown a strong correlation between cluster
metallicity and LMXB incidence, with metal-rich clusters being much more likely
to harbor an LMXB \citep[by a factor of 3 or more;][]{2004ApJ...613..279J,2007ApJ...660.1246S}.
However, this result appears to hold only for bright LMXBs.  When looking at
the low-luminosity LMXB and CV populations in our Galaxy (which comprise the majority
of sources in Figure~\ref{fig:gamma}), a correlation between
source incidence and cluster metallicity is not clearly apparent \citep{2006ApJ...651.1098H}.
In any case, the metallicities of the three overabundant clusters are not significantly
larger than those of the other clusters in Figure~\ref{fig:gamma}, with [Fe/H]
values of -1.30, -1.95, -2.12 for Ter 1, NGC 6397, and NGC 7099, respectively,
with the rest of the clusters ranging in value from -0.34 (NGC 6440) to 
-1.75 (NGC 6093) \citep{1996AJ....112.1487H}.  

Another possibility is that the XRBs in the low-$\Gamma$ clusters
may be mainly primordial, in which case their number should scale with
cluster mass.  However, were this the case, a look at cluster
absolute magnitudes shows that NGC 6121 would have
roughly as many sources as NGC 7099 and more than NGC 6397 and Ter 1, and 
NGC 6366 would have more than Ter 1 \citep{1996AJ....112.1487H}.
Note that the 2--5 sources detected in the low-$\Gamma$ cluster NGC 288 (not included in this 
study), with $\Gamma\approx 0.53$ in the \citet{2003ApJ...591L.131P} normalization, 
are likely primordial \citep{2006ApJ...647.1065K}.

\section{Discussion}\label{sec:disc}

This {\em letter} presents the confluence of three suggestive observational and theoretical 
results into a self-consistent picture.  The first is that of the clusters
that have been observed sufficiently to determine their XRB population, the three
that are core-collapsed are the same three that have a significant X-ray source overabundance
(of a factor of $\sim 2$ to $\sim 20$).
The second is the semi-analytical result derived in this {\em letter} that a cluster in the binary
burning phase for the last few Gyr should have $\sim 5$ times more dynamically formed X-ray sources than if it 
were in the core contraction phase for the same time.  The third is the recently confirmed discrepancy 
between observations and theory for the core radii of Galactic GCs,
which suggests that only the observationally core-collapsed clusters are in the binary-burning phase.
In light of these facts, the conclusion that seems strongly suggested
is that most Galactic GCs are currently still in the core contraction phase, 
while only the $\sim 20\%$ of clusters that are core-collapsed are in the binary
burning phase.  This goes counter to the widely held belief that most clusters 
are currently in the binary burning phase, and complicates the many existing studies that have assumed
cluster core properties that are constant with time.

The implications of this result are manifold.  There are many studies of the dynamical production of 
interesting source populations in clusters which assume core properties that are constant with time.  These
include predictions of the formation of blue stragglers \citep[e.g.,][]{2004ApJ...605L..29M}, the evolution of
the core binary fraction \citep{2005MNRAS.358..572I}, and tidal-capture binaries 
\citep[e.g.,][]{1994ApJ...423..274D}, among others.
Revising the results may be as simple as scaling predicted source numbers,
but may not be so simple for other quantities.

Studies of GC evolution have shown that clusters starting from
very different initial conditions evolve toward a common range of values in many
observable structural parameters in the binary-burning phase, including the
concentration and ratio of core to half-mass radius
\citep{2003ApJ...593..772F,2006MNRAS.368..677H,2007ApJ...658.1047F}.  
Since most clusters may
not be in the binary-burning phase after all, their observed properties are likely
to be more strongly correlated with their initial conditions.  This makes modeling
of clusters a bit more complicated, but on the other hand allows one to more readily deduce 
something about the initial properties of clusters.

Perhaps anticlimactically, our results suggest that the alternative energy sources
recently proposed for supporting GC cores are not
required.  These include the suggestion of IMBHs in {\em many} 
Galactic GCs \citep{2006astro.ph.12040T},
enhanced stellar mass loss from stellar evolution of 
physical collision products \citep{chatterjeeposter}, 
mass segregation of compact remnants in young clusters \citep{2004ApJ...608L..25M}, 
or evaporation of the stellar-mass black hole subsystem in young clusters 
\citep{2007MNRAS.379L..40M}.

Although the picture painted in this {\em letter} is a suggestive one, there
are still several caveats and limitations to our analysis.  In the 
derivation of our refined $\Gamma$ we assume that the core binary fraction 
and abundance of compact objects are constant over the time of integration.
Neither is strictly true, although we expect they will not vary enough
to significantly change the overabundance value we derive.  
We have used only one possible expression for the evolution of the core radius
in core contraction.  While we expect the general behavior to be very similar to
what we have assumed here, more work needs to be done to determine if it is 
universal.  Additionally, our analysis ignores the effect of Galactic tidal stripping on 
cluster mass, which would make a cluster appear overabundant in X-ray
sources, and may be relevant for NGC 6397 \citep{2003ApJ...591L.131P}.
On the observational side, there are some uncertainties in 
evaluating $\Gamma$, which is dependent on quantities that are somewhat 
difficult to measure for core-collapsed clusters.

\acknowledgements

The author thanks T. Fragkos, C. Heinke, N. Ivanova, V. Kalogera, D. Pooley, F. Rasio,
and J. Sepinsky for helpful discussions, D. Pooley in particular
for data and for pointing out the importance of primordial X-ray populations
in low-$\Gamma$ clusters, and the referee P. Edmonds for comments and
suggestions that improved this work.
JMF acknowledges support from Chandra theory grant TM6-7007X and Chandra 
Postdoctoral Fellowship Award PF7-80047.

\bibliographystyle{apj}
\bibliography{apj-jour,main}

\clearpage

\end{document}